\documentclass{article}
\usepackage{arxiv}
\usepackage[utf8]{inputenc}
\usepackage[T1]{fontenc}
\usepackage{graphicx}
\graphicspath{{figs/}}
\usepackage{indentfirst}
\usepackage{csquotes}
\usepackage{hyperref}
\usepackage{xcolor}
\usepackage{paralist}
\usepackage{etoolbox}
\usepackage[numbers]{natbib}
\usepackage{amsmath,amsthm,amssymb}
\usepackage{float}

\hypersetup{ colorlinks=true, linkcolor=black, filecolor=black, urlcolor=black }

\usepackage{lipsum}

\begin{document}
\title{IBM Multilevel Process Mining vs de facto Object-Centric Process Mining approaches}
\author{Alberto Ronzoni \and Anina Antony \and Anjana M R \and Francesca De Leo \and Jesna Jose \and Mattia Freda \and Nandini Narayanankutty \and Rafflesia Khan \and Raji RV \and Thomas Diacci}

\date{\today}
\maketitle

\section{Introduction}
The academic evolution of process mining is moving toward object centric process mining, marking a significant shift in how processes are modeled and analyzed.  IBM has developed its own distinctive approach called Multilevel Process Mining. This paper provides a description of the two approaches and presents a comparative analysis of their respective advantages and limitations. IBM leveraged this comparison to drive the evolution of IBM Process Mining product, creating the new Organizational Mining feature, an innovation that combines the best of the two approaches. Demonstrate the potential of this novel, innovative and distinct methodology with an example.
\section{Traditional Process Mining}
\textbf{Process Mining} is an analytical discipline that integrates data science and process science to \textbf{model}, \textbf{analyze} and \textbf{optimize} processes within businesses, public agencies or other organizations.
\\
\\
A \textbf{traditional process mining} approach begins by \textbf{extracting event data} from information systems, such as enterprise resource planning (ERP) or customer relationship management (CRM) tools. Once extracted, the event data are used to create a process model or process graph of the actual process. Subsequently, the end-to-end process is examined to understand the performance of the processes, revealing bottlenecks and other areas for improvement.
\subsection{Data Preparation}
\begin{figure}
    \centering
    \includegraphics[width=1\linewidth]{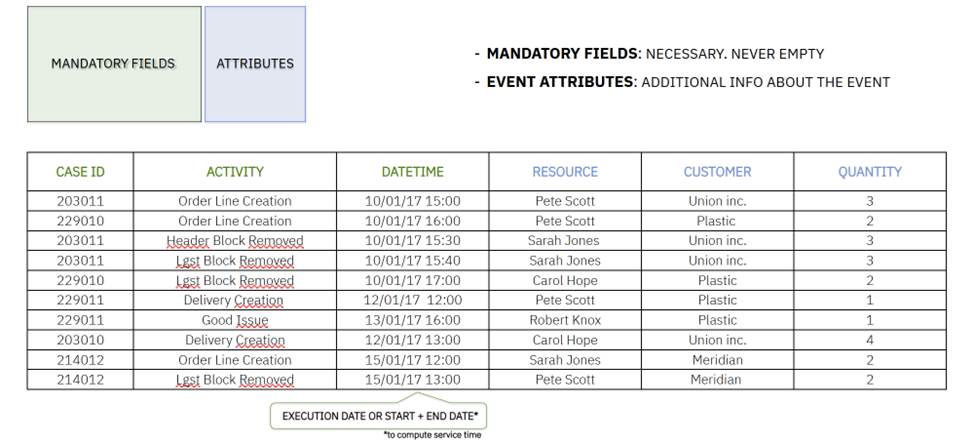}
    \caption{Example of an event log used in process mining, highlighting mandatory fields and additional attributes.}
    \label{fig:Figure_ref_1}
\end{figure}

As previously mentioned, process mining begins with event data extraction. These data are stored in event logs like the one shown in (Figure~\ref{fig:Figure_ref_1}).
Each process instance is called \textbf{case} and can consist of multiple events related to the process. Each case can be viewed as a sequence of events, and an event log can be considered a collection of cases and a case can be seen as a sequence of events.
\\
\\
In an event log, each row corresponds to a specific event inside the business process. An event may consist of several attributes, but only three are mandatory for process mining:
\\
\begin{itemize}
    \item \textbf{A case identifier} (identified by a Case ID);
    \item \textbf{An activity} (the textual description of the event);
    \item \textbf{A timestamp} (when the event occurred).
\end{itemize}
Additional information, such as resources, customers, quantity etc., may also be included but they are not mandatory. These attributes do not directly affect the process discovery, they are utilized - for data drill-down, analytics, other types of analysis (such as root cause and performance analysis).
\\
\\
Multiple rows with the same Case ID identify individual events within the process for that case. In the event log example (Figure~\ref{fig:Figure_ref_1}), the \textbf{Case ID 203011} appears across several rows, each representing a distinct activity such as "\textit{Create Order Line}" or "\textit{Remove Header Block}" and others.
\subsection{Limitations}
Traditional process models assume a single case notion, assuming each case (or process instance) is treated as an isolated entity. In event logs, as shown in the figure (Figure~\ref{fig:Figure_ref_1}), this single case notion is illustrated by the fact that \textbf{one case identifier} (Case ID) per event.
\\
\\
However, this assumption does not reflect the reality of event occurrence in organizational settings, nor does it align with the data storage practices commonly employed in enterprise information systems from which event logs are extracted.
\\
\\
In practice, a single event, such as the creation of a sales order, can involve \textbf{multiple interrelated objects} (customers, sales orders, sales order items, production orders, shipments and invoices) that are all related to each other. \textbf{Traditional process mining} restrict the analysis to a \textbf{single object}, the sales order, and examine how it flows through the sales department and the interaction information between the various departments and other objects involved in the order would be lost. Consequently, critical interaction patterns between departments and the relationships among the various objects involved in the process are obscured and cannot be captured within this limited analytical framework.
\\
\\
Based on the above discussion, the primary limitations of a traditional process mining can be summarized as follows:
\begin{itemize}
    \item \textbf{Data extraction and transformation are laborious and must requires repetition}
    \\
    Traditional process mining requires extracting data from multi-table relational databases of source information systems into a flat event log (e.g., a table) where each event (e.g., a row) refers to a case, an activity, and a timestamp. To obtain different perspectives from the same source data, a new extraction must be performed, which can be a time consuming and resource-intensive process.
    \\
    \item \textbf{Interactions between objects are not captured}
    \\
    When data extracted from a relational database and flattened into a process mining event log where each object instance represents a single case, the interactions between objects defined in the source data are lost. Consequently, any process models generated from the flattened data describe only the lifecycle of an individual object in isolation.
    \\
    \item \textbf{Three-dimensional reality is reduced into two-dimensional event logs and models}
    \\
    In traditional process mining, a separate event log is created for each object type and analyzed separately. Even when these flat event logs are linked through a related cases data model, three-dimensional, object-centric data and models are reduced to two-dimensional, case-centric event logs and process models.
    \\
\end{itemize}
This data flattening leads to two fundamental problems:
\begin{itemize}
    \item \textbf{Data convergence}
    \\
    Consider an ERP system involving concepts such as an ‘Invoice’ and a ‘Payment’.
    \begin{itemize}
        \item If each process instance (case) corresponds to an ’Invoice’;
        \item If a ‘Payment’ is related to several Invoices, then the ‘payment’ event must be duplicated for each process instance in the event log so that the payment appears in association with each invoice.
    \\
    \end{itemize}
    Consequently, the event is duplicated and statistics are inflated resulting in a higher count of payments received than actually existed in reality.
    \\
    \item \textbf{Data Divergence}
    \\
    Consider an ERP system involving concepts such as an ‘Purchase Order’, ‘Pickup’ and ‘Delivery’ events.
    \begin{itemize}
        \item If each process instance (case) corresponds to a ‘Purchase Order’;
        \item To fulfill each purchase order, several pickups and delivery may be needed ;
        \item Although a pickup always corresponds to and precedes a delivery, this information is lost in the event log.
    \\
    \end{itemize}
    As a result, the ‘pickup’ and ’delivery’ events appear interleaved and disconnected, creating incorrect precedence relationships in the process graph. Some of those links may be interpreted as indicating that the same delivery or pickup was performed multiple times (falsely suggesting rework).
\end{itemize}
\section{Multilevel Process Mining}
Multilevel is a historic and distinctive type of process inside IBM Process Mining, created in response to the limitations of traditional process mining to provide users with a unified view of real-world processes involving multiple entities.
\\
\\
In this particular type of process, the case is dynamically identified by discovering the correlation among the business entities that participate in a single process flow. Events from different business entities are mapped in a unified event log according to their mutual relationships, enabling the construction of the actual correlations during event log parsing and thereby creating multilevel case.
\\
\\
Upon completion of process discovery, IBM Process Mining normalizes the set of rows belonging to the same case to obtain the right multiplicity between events, ensuring that throughput time, resource allocation and cost allocations are accurately calculated. One of the most innovative aspects of the distinctive feature is the capability to combine multiple single-entity processes into a unified model.
\\
\\
In statistics computation, IBM Process Mining accounts for the fact that a single event can belong to one or more cases by considering it only once, thereby avoiding both data divergence and data convergence issues.
\newpage
\subsection{Data Preparation}
\begin{figure}
    \centering
    \includegraphics[width=1\linewidth]{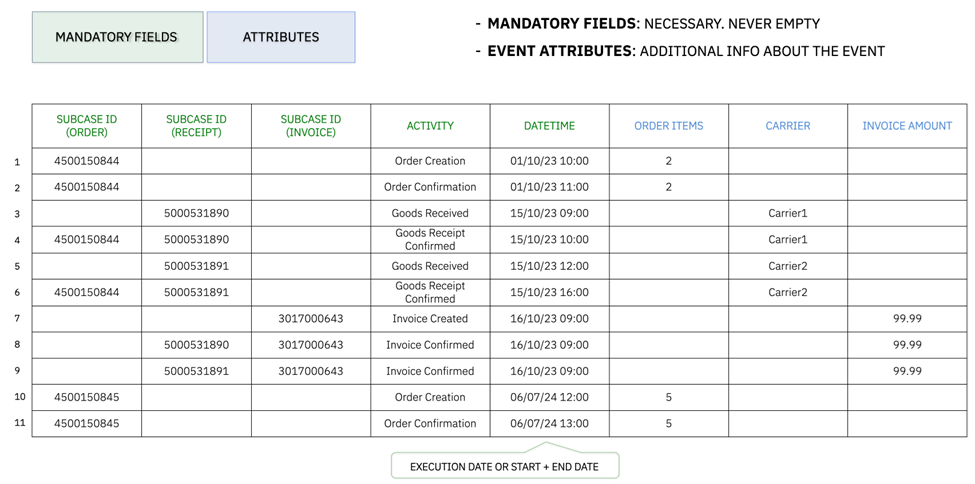}
    \caption{Tabular extraction from a Multilevel event log exemplifying typical relationships between objects (Order, Receipt and Invoice).}
    \label{fig:Figure_ref_2}
\end{figure}

A Multilevel data source must contain an identifying column for each entity (referred to as \textit{ProcessID}) involved in the process. Each \textit{ProcessID} must be chosen such that its value uniquely identifies the corresponding entity. The ordering of the entity sequence must be provided and is important for the correct overall model to be extracted (see \hyperref[sec:Case_Definition]{Case Definition} section).
\\
\\
It is also necessary to define the relationships between these entities, which are connected by bridge activities. A bridge activity does not necessarily have to be the last event of an entity and the and the first event of the next entity. In the example below (Figure~\ref{fig:Figure_ref_2}), if the link between "Invoice” and “Receipt” occurs at the time of “Invoice Confirmed”, that would be the bridge and not the “Invoice Created”.
\\
\\
When a bridge activity needs to link multiple entities (i.e., two “Receipts”) with a single following entity (i.e., “Invoice”), two rows should be defined. The system handles this by creating the dual relationship while considering only one event. Rows 8 and 9 in the portion of the event log shown here (Figure~\ref{fig:Figure_ref_2}) represent this use case.
\\
\\
Canonically, each event should not include more than two populated \textit{ProcessID} columns, as this allows the system to correlate the identifiers and generate the correct case definition. Finally, all non-bridge activities should contain only their respective \textit{ProcessID} (no further links with other entities is required).
\subsection{Case Definition}
\label{sec:Case_Definition}
The mapping order defined for \textit{ProcessIDs} is meaningful, as it should follow the functional or logical flow of the process. This order is used in the identification of the Multilevel case by applying the sequence in reverse order.
\\
\\
Once all events of a specific entity are grouped together, case composition starts from the last entity (“Invoice” in this example); thus, a case will contain only one “Invoice”. Previous entities, linked to invoices through bridge activities, are appended to case. This procedure is executed recursively up to the first mapped \textit{ProcessID}, which is “Order” in this example. If an entity (“Receipt”) is linked to multiple subsequent entities (“Invoices”), it will be included in multiple cases.
\\
\\
Orphan entities may exist at the end of this procedure. For example, an “Order” without a “Receipt” or “Receipt” without an “Invoice” may occur in running cases where the instance of the process is still in progress. To manage this scenario, the above procedure is executed recursively starting from previous entities: “Receipts” and then “Orders”.
\subsection{Process Discovery \& Model Generation}
\begin{figure}
    \centering
    \includegraphics[width=0.6\linewidth]{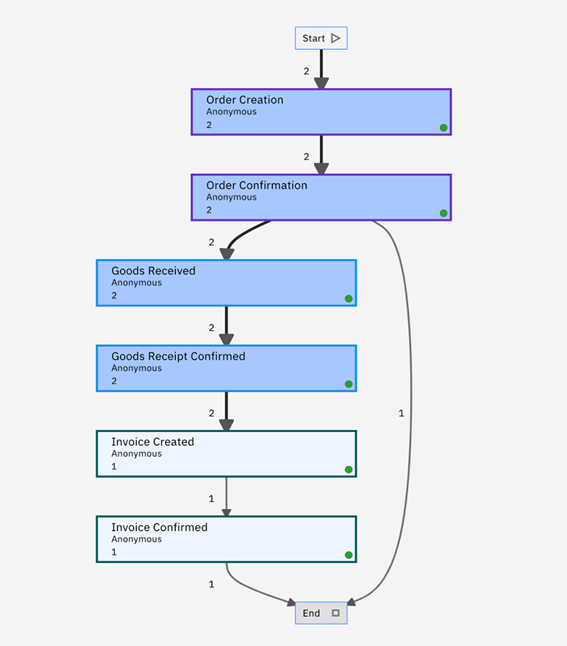}
    \caption{Model of a P2P Multilevel process composed by Order, Receipt and Invoice entities in IBM Process Mining.}
    \label{fig:Figure_ref_3}
\end{figure}

A unique mining algorithm is executed to generate a single unique model (Figure~\ref{fig:Figure_ref_3}). The borderline color of each activity identifies the entity type to which the event belongs.
\\
\\
The multilevel algorithm is computationally expensive compared to traditional mining approaches. Its complexity if affected not only by the number of events but also by ratio between entity relationships within a case.
\\
\\
The mining algorithm supports the presence of multiple entities of the same type within a single case: as shown in the model (Figure~\ref{fig:Figure_ref_2}), no rework is generated between Receipt events (ie., no data divergence), whereas a traditional algorithm would have been generated such artifacts.
\\
\\
For the provided event log, following rules listed above, two cases have been detected. Events 1 through 9 are connected through bridge activities, while events 10 and 11 belong to same entity but are not currently connected to other entities.
\newpage
\subsection{Model Statistics}
\begin{figure}[H]
    \centering
    \includegraphics[width=0.8\linewidth]{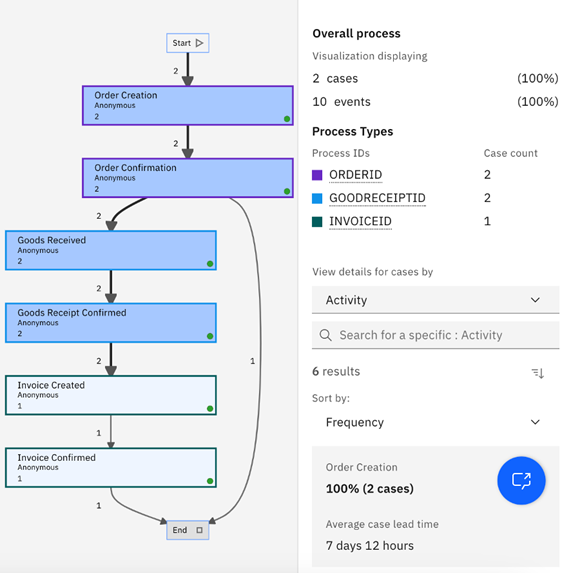}
    \caption{Model and statistics of a P2P Multilevel process composed by Order, Receipt and Invoice entities in IBM Process Mining.}
    \label{fig:Figure_ref_4}
\end{figure}

Upon examination of the event log (Figure~\ref{fig:Figure_ref_2}), it appears to be composed of 11 rows. However, the model displays only 10 events (Figure~\ref{fig:Figure_ref_4}). The discrepancy occurs because rows 8 and 9 were merged in a single event of type “Invoice Confirmed”. As previously explained in the data preparation section, multiple rows were needed to create the connection between two receipts and one invoice. The mining algorithm recognizes this pattern and accounts for it by merging the rows into a single event, thereby avoiding data convergence.
\\
\\
The statistics also display the cardinality of each entity involved in the process. These numbers are accurate because not affected by data convergence. For example, if the same receipt linked to two different invoices (belonging to two different cases) it is counted just once in the statistics. This principle applies to frequency calculations as well as to cost and duration metrics.
\\
\\
The number of cases doesn’t represent a business value metric but rather constitutes the fundamental unit of work in multilevel mining. When filters are applied to the model, an entire case is either retained or discarded. This means that if a user filter for a specific receipt and the case contains multiple receipts, the entire case is remains visible, and the user may observe a receipt count greater than one in the statistics.
\subsection{Reference Model \& Conformance Checking}
\begin{figure}[H]
    \centering
    \includegraphics[width=0.8\linewidth]{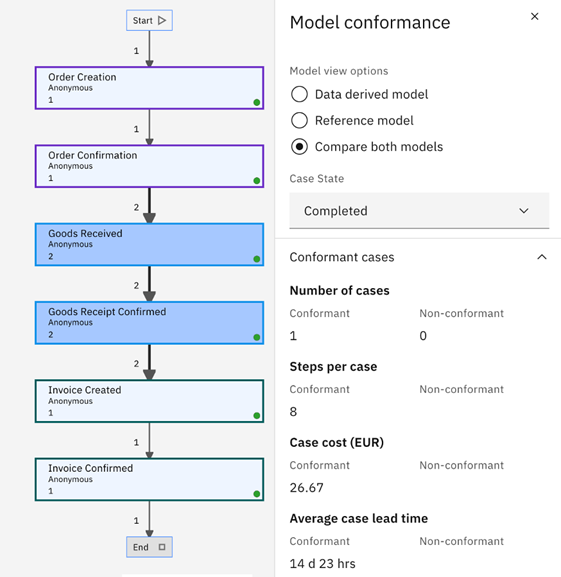}
    \caption{Results of the comparison between data derived and reference model in a P2P Multilevel process composed by Order, Receipt and Invoice entities in IBM Process Mining.}
    \label{fig:Figure_ref_5}
\end{figure}

The approach by which process discovery and model generation are implemented significantly helps conformance checking, making it straightforward and highly similar, at least from the user experience perspective, to traditional process mining conformance checking capability.
\\
\\
The user needs to provide an overall (multi-entity) reference model, following the same methodology as in traditional process mining. Cardinalities of different entity types should not be considered in its definition. The algorithm merely verifies the correct sequence of events, the ratio between entity types is not taken in consideration.
\\
\\
This approach easily supports the discovery of deviations when events of different entity types are mixed. This occurs because conformance checking is not executed at entity level, but rather at the multilevel case level.
\\
\\
This advantage has a limitation, however. If even just one entity in a case is not conformant, the entire case is considered not conformant. For instance, suppose a case contains two orders, but only one is non-conformant. In this scenario, both orders would be flagged as non-conformant because, within the same case, they are considered not independent (see Figure~\ref{fig:Figure_ref_5}).
\subsection{Throughput Time Between Activities}
\begin{figure}
    \centering
    \includegraphics[width=1\linewidth]{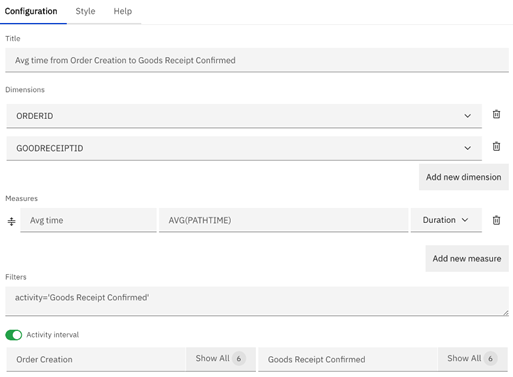}
    \caption{Configuration section for the definition of the computation of throughput time between two activities in a Multilevel process in IBM Process Mining.}
    \label{fig:Figure_ref_6}
\end{figure}

\begin{figure}
    \centering
    \includegraphics[width=1\linewidth]{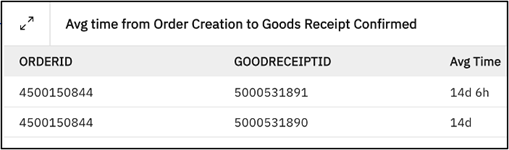}
    \caption{Tabular results of the computation of the average throughput time in a P2P Multilevel process between "Order Creation" and "Good Receipt Confirmed" activities in IBM Process Mining.}
    \label{fig:Figure_ref_7}
\end{figure}

Throughput time can be readily computed selecting any two activities, regardless of whether they belong to different entity types. Multilevel algorithms leverage the case construct to identify linked entities, automatically extract all distinct paths connecting the selected activities.
\\
\\
Consider the computation of average throughput time between “Order Creation” and “Goods Receipt Confirmed” activities (Figure~\ref{fig:Figure_ref_6}). In this scenario, the case construct serves as an enabler rather than a constraint. Although only one case contains both activities, the output (Figure~\ref{fig:Figure_ref_7}) yields two distinct throughput time values. This result reflects the fact that order 4500150844 is associated with two different goods receipts: 5000531891 and 5000531890. The case construct is employed exclusively to identify linked entities, while the enumeration of paths remains independent of case boundaries.
\\
\\
The algorithm identifies  the complete set of paths between the two activities computes the throughput time based on timestamp differentials. When an activity occurs multiple times within the same entity instance, the user can specify whether the first or last occurrence should be considered in the calculation.
\newpage
\section{Object Centric Process Mining}
Object-Centric Process Mining (OCPM) represents an advanced approach to process mining that addresses the inherent complexity of real-world business operations by modeling events as interactions among multiple interconnected objects. In contrast to conventional process mining techniques, which adopt a single-case perspective, OCPM enables a more holistic, comprehensive and integrated representation of business processes.
\\
\\
At its core, Object-Centric Process Mining recognizes that business activities typically involve multiple entities, such as orders, invoices, shipments, and customers, all interacting within the same shared workflow. Rather than assigning each event to a single case identifier, as in traditional process mining, OCPM associates events with multiple objects simultaneously. This approach accurately reflects the true structure of business operations and enables a richer, more accurate representation of workflows.
\\
\\
OCPM enables organizations to analyze how temporal interactions among distinct business objects, uncover hidden dependencies, and detect inefficiencies or bottlenecks that would otherwise remain undetected under traditional analytical approaches. This approach is particularly valuable for businesses with complex, interdependent processes, including supply chain management, financial transaction networks, and enterprise resource planning (ERP) systems.
\\
\\
One of the key advantages of OCPM is its flexibility. Once data from various source systems is transformed into an object-centric format, it can be analyzed from multiple perspectives without requiring additional data extraction or transformation operations. This allows businesses to dynamically explore different aspects of their processes, thereby generating deeper insights and supporting decision-making.
\\
\\
With its ability to model real-world business processes more accurately than traditional methods, Object-Centric Process Mining is becoming an essential tool for organizations seeking to optimize their operational efficiency, enhance process transparency, and enable data-driven strategic decision-making. 
\\
\\
Throughout this work, we make references to the fundamental concepts of Object-Centric Process Mining and draw insights \citep{vanderAalst2023fabric,vanderAalst2023introduction}.
\subsection{Object Centric Event Data-Key Components}
Object-Centric Event Data (OCED) is fundamental to understanding and analyzing business processes that involve multiple interrelated objects or entities, often captured in modern process mining. Unlike traditional event logs that track events in isolation or based on a single case notion, OCED enables the representation of complex relationships between various objects involved in process execution. 
\\
\begin{itemize}
    \item \textbf{Events}: An event represents the execution of a specific activity and serves as the fundamental unit in an event log. It is characterized by a timestamp, indicating when the event occurred, the objects involved in the event, and the values of activity-specific attributes;
    \item \textbf{Objects}: Objects are uniquely identifiable instances of a particular object type. Each object can have attributes whose values may change over time, reflecting the dynamic nature of the object's state throughout the process;
    \item \textbf{Relationships}:
    \begin{itemize}
        \item \textbf{Event-to-Object Relationships}: Every event is linked with at least one object, indicating the objects directly involved in that event;
        \item \textbf{Object-to-Object Relationships}: Objects can also be related to one another without necessarily sharing an event. These relationships define how objects are interlinked within the process, providing a comprehensive view of their interactions.
    \end{itemize}
    \item \textbf{Activities}: Activities define the type of event executed, such as "register customer order." Each activity can have various associated attributes, including the location where the activity took place or the performer of the action;
    \item \textbf{Object Types}: Each object belongs to a specific object type, which defines a set of attributes for the object. These attributes may vary between different objects or evolve over time for the same object, capturing changes in the object's state;
    \item \textbf{Qualifiers}: Relationships in OCED can be qualified to provide additional context. For instance, an object's role in an event, such as identifying the actor responsible for performing the activity can be explicitly distinguished, adding depth and clarity to the event log.
\\
\end{itemize}
This structured approach to capturing events and relationships enables a more detailed and accurate analysis of how different objects and activities interconnect in complex business processes.
\begin{figure}
    \centering
    \includegraphics[width=0.8\linewidth]{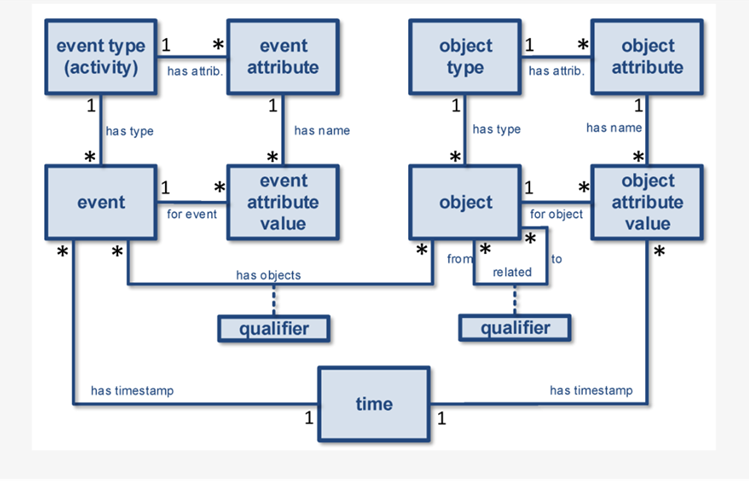}
    \caption{Illustrates the Object-Centric Event Data Meta-Model (OCED-MM): which defines the core concepts and relationships underlying object-centric event data.}
    \label{fig:Figure_ref_8}
\end{figure}
\\
The object-centric event data meta-model (OCED-MM) (Figure~\ref{fig:Figure_ref_8}). The classes (rectangles) introduce the main concepts (events, objects, attributes, etc.). The relationships are annotated with cardinality (* denotes zero or more) e.g., an event may refer to any number of events, an event has precisely one timestamp, and an event has precisely one event type.
\subsection{Relationships}
Understanding relationships in Object-Centric Process Mining (OCPM) is essential for analyzing and optimizing business processes that involve multiple interacting entities. OCPM incorporates two main primary relationship types: Event-to-Object and Object-to-Object, each providing valuable insights on process dynamics.
\\
\\
\textbf{Event-to-Object Relationships}:
\\
In OCPM, the event-to-object relationship is essential for tracking the object progression through a process. Events represent the executed actions or activities , while objects are the entities affected by these actions. This relationship provides a clear trace of each object's journey, such as an order progressing from creation through shipment.
\begin{itemize}
    \item \textbf{Purpose}: These relationships help outline the sequence and timing of events associated with each object. This enables the identification of potential delays, inefficiencies, or bottlenecks in the process, which is critical for performance optimization;
    \item \textbf{Example}: Consider an order object that undergoes various events like "Order Received," "Order Processed," and "Order Shipped." By analyzing these events, enables identification of delays in stages and understand the underlying reasons;
    \item \textbf{Benefits}: The detailed view offered by event-to-object relationships allows for a comprehensive understanding of how events influence an object's lifecycle. This perspective makes it easier to pinpoint areas where intervention may be necessary to maintain process efficiency and achieve desired performance levels.
\\
\end{itemize}
\textbf{Object-to-Object relations}:
\\
Object-to-object relationships, on the other hand, illustrate how different entities within a process are connected, either through direct associations or intermediate linkages. These relationships capture dependencies and interactions essential for ensuring seamless integration of various process components.
\begin{itemize}
    \item \textbf{Types of Relationships}:
    \begin{itemize}
        \item \textbf{One-to-One}: A single order linked to a single invoice;
        \item \textbf{One-to-Many}: An order associated with multiple shipments;
        \item \textbf{Many-to-Many}: Multiple orders associated with multiple invoices.
    \end{itemize}
    \item \textbf{Purpose}: Understanding these relationships clarifies information flows between entities and reveals dependencies that may influence overall process performance. Understanding these structures helps in detecting potential bottlenecks and ensures efficient inter-entity communication and data exchange;
    \item \textbf{Example}: An order may be associated with multiple shipments, each requires coordination to ensure timely order fulfillment. Analyzing these relationships helps identify the potential delays, uncovering paths affecting the order's completion;
    \item \textbf{Benefits}: Examining object-to-object interactions provides a holistic view of the entire process. It aids in understanding of interdependencies that impact performance and enables improved coordination and resource management to achieve desired outcomes.
\\
\end{itemize}
By analyzing both event-to-object and object-to-object relationships, OCPM provides a comprehensive insight into processes behavior and optimization opportunities. This enables better decision-making and more effective process optimization.
\newpage
\subsection{Data Preparation}
\begin{figure}
    \centering
    \includegraphics[width=1\linewidth]{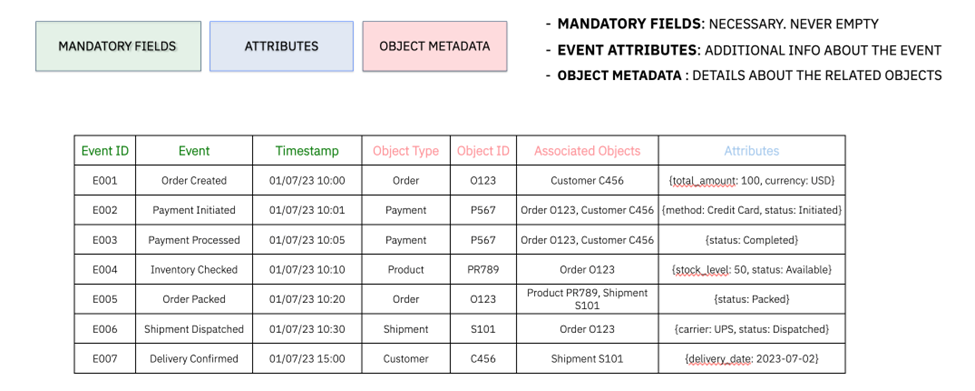}
    \caption{Event Log Structure for Object-Centric Process Mining (OCPM): illustrates an example of an object-centric event log containing mandatory fields, event attributes, and object metadata.}
    \label{fig:Figure_ref_9}
\end{figure}

\begin{figure}
    \centering
    \includegraphics[width=1\linewidth]{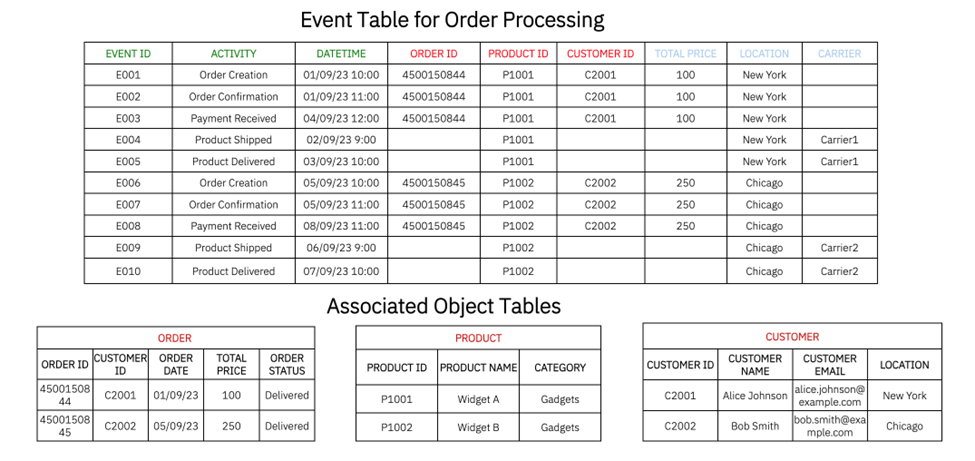}
    \caption{Event and Object Tables for Order Processing: shows the separation of event and object data into related tables, linking activities with entities such as orders, products, and customers to enable object-centric process analysis.}
    \label{fig:Figure_ref_10}
\end{figure}

Object-Centric Event Log (OCEL) serves as a standardized data format specifically designed for process mining applications, optimized to capture the interactions between processes and the objects involved in these processes.
\begin{enumerate}
    \item \textbf{Identification of Relevant Objects and Events}
    \begin{itemize}
        \item \textbf{Objects}: Identify the entities interacting with the process, which may include physical objects (e.g., products, machines), abstract entities (e.g., cases, orders), or human actors;
        \item \textbf{Events}: Events denote the actions or occurrences within the process, with each event explicitly associated with one or more specific objects.
    \end{itemize}
    \item \textbf{Structure the data}
    \\
    Organizing the data systematically is essential to support efficient process mining and analysis.
    \begin{itemize}
        \item \textbf{Event Log Structure}: Construct a well-defined table or data structure where each row represents an event. This structure should capture all relevant details of each event in a clear and accessible manner;
        \item \textbf{Event Attributes}: Each event record must include key attributes to facilitate comprehensive analysis:
        \begin{itemize}
            \item \textbf{Event ID}: A unique identifier assigned to each event to distinguishing each event instance;
            \item \textbf{Timestamp}: The exact date and time when the event occurred, enabling analysis of the sequence and timing of activities;
            \item \textbf{Activity}: The name or type of action performed, such as "\textit{Order Received}" or "\textit{Product Shipped}";
            \item \textbf{Object ID}: The identifier of the object involved in the event, linking the event to its corresponding entity.
        \end{itemize}
    \end{itemize}
    \item \textbf{Establish object relationships}
    \\
    Understanding how objects interact and transition through different stages of the process is essential for comprehensive process analysis.
    \begin{itemize}
        \item \textbf{Object Lifecycle}: Define the possible states or stages that an object can traverse during process execution. For example, an order might progress through stages such as "created," "processed," and "shipped";
        \item \textbf{Object Interactions}: Analysis of inter-objects relationship reveals how distinct entities interact with one another. These interactions could be:
        \begin{itemize}
            \item \textbf{Creation Links}: Where one object is responsible for creating another;
            \item \textbf{Hierarchical Relationships}: Where objects are organized in a parent-child structure, such as a shipment containing multiple products. 
        \end{itemize}
    \end{itemize}
    \item \textbf{Ensure Data quality}
    \\
    Maintaining high-quality data is essential to derive meaningful and accurate insights from process mining analysis.
    \begin{itemize}
        \item \textbf{Completeness}: The event log must capture all necessary events and object information. Missing or incomplete data can lead to an partial process representations that can compromise the process mining analysis;
        \item \textbf{Consistency}: Ensure that the data is free from contradictions or inconsistencies. For example, timestamps should follow a logical order, and object states should progress as expected;
        \item \textbf{Accuracy}: Regularly validate the correctness of the data. Errors or inaccuracies can compromise the reliability of the analysis and the validity of derived conclusions;
        \item \textbf{Timeliness}: Keep the data up-to-date to ensure that it accurately reflects the current state of the process. Timely data ensures the relevance and actionable insights.
    \end{itemize}
\end{enumerate}
Adherence to these guidelines for objects and events identification, data structuring, relationship definition, and data quality assurance enables a robust Object-Centric Event Log that serves as a reliable foundation for process mining and analysis. This comprehensive approach enables a more accurate and insightful examination of complex business processes, facilitating data-driven optimization and continuous improvement initiatives.
\\
\\
Below (Figure~\ref{fig:Figure_ref_9}, Figure~\ref{fig:Figure_ref_10}) presents visual representations illustrating the data preparation for Object-Centric Process Mining (OCPM). These visualizations demonstrate the structured data, relationships, and key components that constitue the foundation for analyzing complex business processes through OCPM framework.
\newpage
\subsection{Eventlog Data \& Activity Characteristics}
\begin{figure}[H]
    \centering
    \includegraphics[width=1\linewidth]{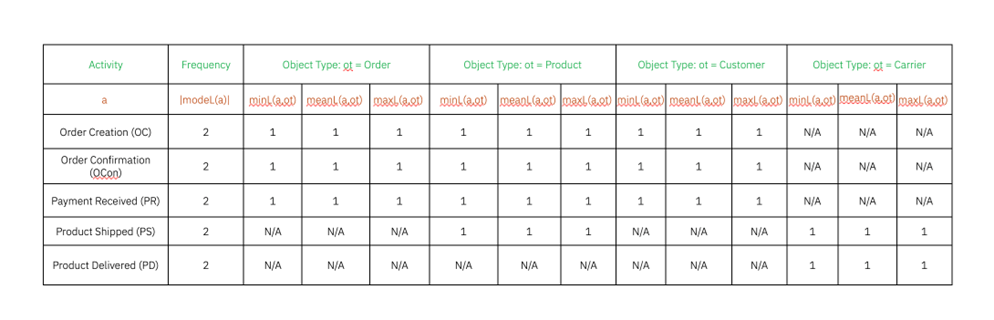}
    \caption{Activity Characteristics in Object-Centric Process Mining (OCPM): depicts the interaction between activities and object types—Orders, Products, Customers, and Carriers—in an order processing system.}
    \label{fig:Figure_ref_11}
\end{figure}

In Object-Centric Process Mining (OCPM), \textbf{activity characteristics} refer to the specific properties and behavioral patterns of activities within a process, particularly how distinct types of objects interact with those activities. These characteristics provide a deeper insights into process execution, and help identify patterns, bottlenecks, and opportunities for optimization.
\\
\\
The below image (Figure~\ref{fig:Figure_ref_11}) provides a detailed representation of the activity characteristics in Object-Centric Process Mining (OCPM) as applied to an order processing system. These visuals illustrate the interactions between activities and various object types, including Orders, Products, Customers, and Carriers, throughout the process.
\\
\\
The data reflects the frequency, execution modes, and object type interactions for different activities such as "Order Creation", "Order Confirmation", "Payment Received", "Product Shipped", and "Product Delivered". For each activity, the minimum, mean, and maximum number of objects involved are presented, providing a comprehensive view into the complexity and variability of object involvement in the process.
\\
\\
Key analytical dimensions such as Execution Mode, Activity Occurrence, and Activity-Object Type Combinations are visualized, along with the analysis of object counts across different activities. This structure enables for an in-depth examination of object types interactions during process execution, facilitating pattern identification, bottleneck detection, and process optimization opportunity.
For an event log L=(E,O,\#,R) with activities A=act(L) and object types OT=types(L), we define the following concepts and functions on \textbf{Order Processing Event Log}.
\begin{itemize}
    \item \textbf{Execution Mode}: An execution mode \[m \in B(OT)\] is a multiset of object types and characterizes the types of objects and their count involved in an event. (e.g., order, product, customer). In this case, the relevant object types are Order, Product, Customer.
    \\
    In a simple way Execution modes describe the types and quantities of objects involved in an activity. For example, an activity like "Order Creation" may involve an Order, a Product, and a Customer. The execution mode for this activity would be a multiset representing these objects and their counts;
    \item \textbf{Activity Occurrence}: An activity occurrence \[(a,m) \in A×B(OT)\] refers to an activity a executed in mode m. In a simple way this refers to the execution of an activity involving specific objects. An activity occurrence is a combination of an activity and its corresponding execution mode.  For example, the activity "\textit{Order Creation}" could involve one Order, one Product, and one Customer. Another activity like "\textit{Product Shipped}" may involve only a Product and Carrier;
    \item \textbf{Activity-Object Type Combination} aot(L): This is the set of all activity-object-type combinations in event log L . For example, "\textit{Order Creation}" may involve an Order, Product, and Customer objects;
    \item \textbf{Occurrence of Activity in Log} occ(L): This is a multiset of all activity occurrences in L. For example, "\textit{Order Creation}" could occur multiple times, each involving different objects like different Orders, Products, or Customers;
    \item \textbf{Execution Modes for Activity} mode L(a): This represents the multiset of execution modes for activity a. For instance, the "\textit{Order Creation}" activity may occur in different modes depending on the number of objects involved (e.g., one Order, one Product, and one Customer);
    \item \textbf{Minimum Number of Object Types per Activity} minL(a,ot): The minimum number of objects of type \[ot \in OT\] involved when executing activity a. For example, the minimum number of Orders involved in the "\textit{Order Creation}" activity is 1;
    \item \textbf{Maximum Number of Object Types per Activity} maxL(a,ot) : The maximum number of objects of type \[ot \in OT\] involved when executing activity a. For example, in the "\textit{Product Shipped}" activity, the maximum number of Products is 1;
    \item \textbf{Mean Number of Object Types per Activity} meanL(a,ot): This is the mean number of objects of type \[ot \in OT\] involved when executing activity a. For instance, the mean number of Products involved in the Order Confirmation activity could be calculated based on how often different products are confirmed for orders.
    \\
    \\
\end{itemize}
This formalization allows us to analyze how different types of objects are involved in the activities of your order processing system, such as how many products are typically shipped in a single event or how often customers place multiple orders in the same activity. These insights can be leveraged for performance analysis, process optimization, or detecting anomalies in the process flow.
\subsection{Process Discovery \& Model Generation}
In object-centric process mining, \textbf{process discovery} involves creating models that capture the interactions and dependencies among multiple objects or entities within a unified process representation. Unlike traditional case-centric process discovery, where models focus on events sequence for a single case (e.g., a customer order), object-centric process mining discovers processes that represent how multiple entity types interact within complex workflows.
\\
\\
In object-centric process mining, the event log associates events with multiple objects rather than a singular case identifier. Each event can reference multiple entities, such as an order, product, and customer, enabling the log to represent complex multi-object interactions.
\\
\\
\textbf{Object-Centric Model Types}:
\begin{itemize}
    \item \textbf{Object-Centric Petri Nets (OCPNs)}: These models extend traditional Petri nets to include multiple objects, where tokens represent different types of entities, and transitions represent activities involving these entities;
    \item \textbf{Object-Centric Causal Nets (OCCNs)}: OCCNs use a causal net structure that highlights cause-and-effect relationships between activities across objects. Arcs between activities represent dependencies and interactions among different object types;
    \item \textbf{Object-Centric BPMN Models}: These models adapt the Business Process Model and Notation (BPMN) standard to incorporate multiple objects. Swim lanes or pools can represent distinct entity types, with interactions displayed as message flows or connectors.
    \\
\end{itemize}
\textbf{Key Components of Model Generation}:
\begin{itemize}
    \item \textbf{Object Identification}: Identifying the key objects or entities participating in the process;
    \item \textbf{Lifecycle Modeling}: Definition of states or stages that objects go through during their lifecycle and identifying the transitions between states;
    \item \textbf{Interaction Modeling}: Capturing the interactions between objects and how they influence each other's behavior also Modeling dependencies and constraints;
    \item \textbf{Event-Based Approach}: Using event logs to capture the sequence of events related to objects. Analyzing these events to understand object lifecycles and interactions;
    \item \textbf{Model Representation}: Choosing a suitable modeling language or frameworks to represent the object-centric model including Class diagrams, State diagrams, Activity diagrams, and Petri nets;
    \item \textbf{Model Refinement and Validation}: Iteratively refining the model based on feedback and additional data, with Validation against real-world process data to ensure model accuracy.
    \\
\end{itemize}
\textbf{Handling Overlapping Flat Models in Process Discovery and Model Generation}:
\begin{itemize}
    \item \textbf{Hierarchical Object Models}: Breaking down complex objects into simpler components to reduce overlap, using hierarchical structures to represent inter-objects relationships;
    \item \textbf{Event-Based Approaches}: Using event logs to capture the object-related event sequences, with subsequent analysis to identify patterns and relationships;
    \item \textbf{Semantic Process Mining}: Using ontologies to define the domain-specific concepts and relationships between objects. Providing a formal analytical framework for addressing overlapping models;
    \item \textbf{Hybrid Approaches}: Integration of multiple techniques to address specific modelling challenges and optimize discovery outcomes. 
\end{itemize}
\subsection{Model Statistics}
Model statistics in Object-Centric Process Mining provide analytical insights into the behavioral pattern and interactions of different objects within a process. These metrics facilitate understanding of object roles and their influence on overall process dynamics. Key metrics include object selection and leading objects, which are essential for targeted analysis and process optimization.
\\
\\
\textbf{Object Selection}:
\\
Object selection involves choosing specific objects or instances to focus on within the process mining analysis. By filtering the data, analysts can concentrate on particular objects of interest, such as orders, products, or customers, to analyze their specific roles and impacts. For example, when analyzing the performance of a particular product, selection focuses exclusively on the events and interactions related to that product. This approach enables a detailed examination of how the product affects the process and its interactions with other objects. Object selection refines the analytical scope of the process mining model, facilitating pattern identification and issue detection related to the selected objects. This enhances the ability to conduct targeted analysis and derive actionable insights for process improvement.
\\
\\
\textbf{Leading Objects}:
\\
Leading objects are entities that significantly drive or influence the process execution. These are typically objects that initiate key events or have a substantial impact on the process flow, such as high-priority support tickets or frequently ordered products. In a ticketing system, for instance, identifying leading objects might involve recognizing tickets that are frequently escalated or that have long resolution times. These tickets are critical as they often reveal bottlenecks or areas requiring intervention. Focusing on leading objects facilities understanding of the key drivers of process performance. By addressing issues related to these objects, organizations can implement targeted improvements and enhance the overall process efficiency and effectiveness.
\subsection{Conformance Checking}
Conformance checking is the process of verifying whether the actual sequences of activities and interactions between objects align with a predefined model or set of rules. This ensures that observed processes conform to the expected behavioral standards as defined by organizational protocols or process models.
\\
\\
\textbf{Multi-Object activity sequences}:
\\
OCPM involves analyzing sequences of activities that involve multiple objects (e.g., products, orders, customers). Conformance checking must account for these complex multi-object interactions, ensuring that the sequence and flow of activities across different objects adhere to the specified process model.
\\
\\
\textbf{Allowed Activity occurrences}:
\\
OCPM allows the validation of whether specific activities or events associated with various objects occur as permitted or expected within the process. This involves verifying whether each object's activities conform to the allowed sequences defined in the process model specification.
\\
\\
\textbf{Cross-Object relationships}:
\\
Conformance checking in OCPM must also consider the relationships and dependencies between objects. For example, when two objects are required to interact in a prescribed way at a specific process stage, the conformance verification confirms whether this interaction occurs as specified.
\\
\\
\textbf{Deviations and Anomaly detection}:
\\
The conformance checking process involves identifying deviations or anomalies where the observed behavior diverges from the expected model. These deviations may indicate process inefficiencies, execution errors, or opportunities for process improvement.
\\
\\
\textbf{Handling Many-to-Many relationships}:
\\
In OCPM, it's common to have many-to-many relationships between events and objects. Conformance checking must effectively manage these complex data structures to ensure accurate verification of allowed activity occurrences across all related objects.
\\
\\
\textbf{Adaptive Conformance Checking}:
\\
Given the dynamic nature of processes involving multiple objects, the models used for conformance checking in OCPM need to be flexible. These models should accommodate process variations while maintaining a robust framework for detecting significant deviations from expected behavior.
\subsection{Throughput Time Between Activities}
\begin{enumerate}
    \item \textbf{Object-Level throughput time}:
    \begin{itemize}
        \item \textbf{Object-centric activities}: Define activities that are specific to a particular object, such as "\textit{Process Order}" or "\textit{Ship Product}";
        \item \textbf{Object-level duration}: Calculate the time it takes for an object to complete a specific activity;
        \item \textbf{Object-level throughput time}: Determine the overall time taken by an object to progress through its lifecycle.
    \end{itemize}
    \item \textbf{Activity-level throughput time}:
    \begin{itemize}
        \item \textbf{Activity-centric activities}: Define activities that are common across multiple objects, such as "\textit{Receive Payment}" or "\textit{Assign to Resource}";
        \item \textbf{Activity-level duration}: Calculate the time it takes to complete an activity, regardless of the specific object involved;
        \item \textbf{Activity-level throughput time}: Determine the average time taken to complete an activity across all objects.
    \end{itemize}
    \item \textbf{Linked Entities and Relationships}:
    \begin{itemize}
        \item \textbf{Linked entities}: Identify relationships between objects that influence throughput time, such as "order" and "product," or "customer" and "case.";
        \item \textbf{Relationship-based throughput time}: Calculate the time it takes for an object to progress through a specific relationship or interaction with another object.
    \end{itemize}
\end{enumerate}
\newpage
\section{Object Centric vs Multilevel}
Do a comparison between the two to highlight the strengths and weaknesses of both solutions.
\subsection{Similarities}
\begin{itemize}
    \item \textbf{Focus on Granularity and Complexity}:
    \begin{itemize}
        \item \textbf{Object centric process mining}:
        Object-centric process mining delves into the intricate relationships between different objects within a process, providing a detailed view of how they interact and influence each other.
        \item \textbf{Multilevel process mining}:
        Multilevel process mining explores processes at multiple levels of abstraction, from high-level, overarching activities to granular, low-level tasks.
    \end{itemize}
    \item \textbf{Beyond Traditional Case-Based Analysis}:
    \begin{itemize}
        \item \textbf{Object centric process mining}:
        Object-centric process mining moves beyond the limitations of traditional process mining, which often focuses on a single case or instance. OCPM considers multiple objects and their interdependencies.
        \item \textbf{Multilevel process mining}:
        Multilevel process mining offers a more holistic perspective by analyzing processes at different levels of detail, providing a deeper understanding of their structure and dynamics.
    \end{itemize}
    \item \textbf{Enhanced Insights into Process Dynamics}:
    \begin{itemize}
        \item \textbf{Object centric process mining}:
        Object-centric process mining uncovers hidden relationships and dependencies between objects, enabling organizations to identify bottlenecks, inefficiencies, and opportunities for optimization.
        \item \textbf{Multilevel process mining}:
        Multilevel process mining provides a broader view of the process landscape, allowing for a better understanding of how high-level goals and strategies are translated into operational activities.
    \end{itemize}
    \item \textbf{Support for Decision Making}:
    \begin{itemize}
        \item \textbf{Object centric process mining}:
        Object-centric process mining offers valuable insights for process improvement initiatives, enabling organizations to make data-driven decisions and optimize their operation.
        \item \textbf{Multilevel process mining}:
        Multilevel process mining provides a framework for aligning strategic objectives with operational execution, supporting effective decision-making at all levels of the organization.
    \end{itemize}
\end{itemize}
\subsection{Advantages of Multilevel Process Mining}
\begin{itemize}
    \item \textbf{Focus on End-to-End Processes}:
    Multilevel process mining provides a comprehensive view of processes spanning multiple entities or organizational units. By analyzing the entire lifecycle of a process, it offers valuable insights into how various components interact and contribute to overall operational efficiency.
    \item \textbf{Cross-Entity Conformance Checking}:
    One of the key advantages of Multilevel process mining is its capability to verify whether actual process execution align with expected or modeled behaviors across multiple entities. This ensures process compliance with intended specifications, thereby improving compliance and operational accuracy.
    \item \textbf{Cross-Entity Event Sequencing}:
    Multilevel process mining captures and analyzes sequences of events across different entities, such as departments or organizational units. Instead of focusing solely on events within a single entity, it considers how events relate and interact across multiple entities, enabling a more interconnected and insightful process analysis.
    \item \textbf{Holistic Optimization}:
    By analyzing processes that spanning multiple entities, organizations can identify inefficiencies and bottlenecks affecting the entire process chain. This broader perspective enables more effective optimization strategies that improve overall process performance rather than just isolated process segments.
\end{itemize}
\subsection{Limitations of Multilevel Process Mining}
\begin{itemize}
    \item \textbf{Limitation of "Case" for Analysis}:
    Traditional process mining relies on the concept of a "case" to structure analysis. However, in Multilevel process mining, this concept may not fully capture the complexity of multi-entity interactions. Consequently, the level of granularity in insights may be reduced compared to traditional case-based analysis.
    \item \textbf{Scalability Challenges}:
    As the number of entities and inter-entity relationships increases, the computational complexity of Multilevel process mining grows exponentially. Managing and analyzing large volumes of interconnected data can be challenging, potentially limiting its scalability for large-scale or highly intricate processes environments.
    \item \textbf{Limited Applicability to Subprocess Analysis}:
    While multilevel process mining is highly effective for end-to-end process analysis, it may not be as useful for examining subprocesses or smaller, more granular components of a larger process flow. This can limit its applicability in scenarios where organizations need detailed insights at a micro-level.
    \item \textbf{Event Log Preparation Complexity with Bridge Activities}:
    Preparing event logs for Multilevel process mining can be complex, particularly when addressing bridge activities that span multiple entities. Ensuring proper data integration and consistency across different process levels may require significant effort and expertise.
    \item \textbf{Limited Industry Adoption}:
    Despite its potential benefits, Multilevel process mining has not yet achieved widespread industry adoption. Factors such as implementation complexity, resource requirements, and perceived applicability may contribute to its limited acceptance in the industry.
    \item \textbf{Complexity in Interpretation and Actionability}:
    The insights generated by Multilevel process mining can be highly complex and challenging to interpret, especially for stakeholders lacking process mining expertise. This complexity can make it difficult for organizations translate findings into clear, actionable improvements without significant analytical expertise.
\end{itemize}
\subsection{Advantages of Object-Centric Process Mining}
\begin{itemize}
    \item \textbf{Overall Model Overlapping Flat Event Logs}:
    Object-centric process mining (OCPM) constructs overall process model by analyzing flat event logs. This enables a detailed understanding of how each individual objects (cases) interact with the process temporally, facilitating comprehensive process insights.
    \item \textbf{Business Intelligence (BI) Performance Analytics}:
    OCPM provides analytical insights into the behavior of individual cases within processes. By supporting data-driven decision-making, it enables organizations to identify bottlenecks, inefficiencies, and process optimization opportunities.
    \item \textbf{OCEL Standard (Object-Centric Event Language) Compliance}:
    OCPM utilizes the OCEL standard for event definition and modeling object-centric process mining. This promotes consistency and interoperability in OCPM implementations, ensuring seamless integration with various tools and methodologies.
    \item \textbf{Scalability}:
    Designed to efficiently handle large volumes of event data, OCPM is suitable for analyzing processes with numerous cases and events. Its scalable architecture ensures its applicability to complex and data-intensive workflows.
    \item \textbf{Support for Subprocesses}:
    Unlike traditional process mining approaches, OCPM supports the analysis of subprocesses within larger processes. This provides more granular insights into specific process segments, improving process optimization efforts.
    \item \textbf{Support for Complex Event Relationships}:
    OCPM enables the analysis of complex event relationships, including parallel and interdependent events. Traditional process mining methods often struggle to capture these intricate dependencies, but OCPM provides a more accurate and comprehensive representation.
    \item \textbf{Cross-Organizational Process Analysis}:
    OCPM is well-suited for processes that span multiple organizations or departments. By enabling the analysis of cross-organizational workflows, it helps identify bottlenecks and inefficiencies across different units, fostering better collaboration and process efficiency.
    \item \textbf{Process Flexibility Analysis}:
    OCPM can accommodate and analyze process variations that involve different types of objects. This flexibility allows organizations to study and understand dynamic processes where multiple entities interact, making it easier to implement adaptive process strategies.
\end{itemize}
\subsection{Limitations of Object-Centric Process Mining}
\begin{itemize}
    \item \textbf{High Computational Complexity}:
    OCPM can be computationally intensive due to the need to analyze and correlate large data across volumes across multiple objects. This results in increased processing time and elevated resource requirements, making it challenging to implement in environments with limited computational capacity.
    \item \textbf{Data Preparation Complexity for Many-to-Many Relations}:
    Handling many-to-many relationships between events and objects in event logs can be complex. Proper data preparation is essential, but it is often time-consuming and requires significant expertise to ensure data accuracy and completeness.
    \item \textbf{Limited Full OCPM Use Cases (Mainly in Supply Chain)}:
    Despite its potential, OCPM's full-scale adoption remains limited. It is primarily used in supply chain management, which restricts its general applicability. The scarcityof diverse industry use cases hinders its widespread implementation.
    \item \textbf{Partial Cross-Object Process Mining Insights}:
    A key challenge of OCPM is the lack of standardized Key Performance Indicators (KPIs) specifically designed for cross-object process mining scenarios. This limitation complicates performance benchmarking and compare performance across different processes or industries using OCPM.
\end{itemize}
\section{Organization Mining}
Following the comparative analysis, the product evolution strategy integrated  the optimal characteristics of both Multilevel and OCPM paradigm into a new feature in IBM Process Mining: Organization Mining.
\\
\\
The key technical features of this new approach are the following:
\begin{itemize}
    \item Support for the NextGen mining engine already available in IBM Process Mining, based on a relational database architecture.
    \item Support for large events volume and significantly faster speeds compared to the legacy engine.
    \item Take advantage of SQL's flexibility with structured data across multiple linked tables and its interoperability with external systems.
\end{itemize}
While IBM Process Mining's Nextgen engine provides the technical capabilities to address the computational limitations of the Multilevel approach, the OCPM structure that adds significant functional enhancements to these advantages.
\\
\\
The primary process phase to be impacted and substantially simplified by this approach is data preparation. By eliminating the need to insert all relevant data into the flat event log structure, data can now be organized into separate, interconnected tables (OCPM objects). This ensures that the Organization Mining data structure closely resembles that of the source systems from which process data is extracted, potentially leading to a truly significant simplification of data preparation.
\\
\\
The ability to combine multiple event logs associated with interconnected objects into a single unified view enhances the analytical capabilities for individual processes, enabling more comprehensive organizational-level analysis. This allows analysts to understand how interconnected processes influence one another and their integration points. Here, the advantages of the unified Multilevel model combine with those of a common OCPM view, resulting in a unique and distinctive feature of IBM Process Mining capable of unlocking powerful insights for customers.
\subsection{Data Preparation}
\subsubsection{Challenges in Traditional Process Mining \& the OCPM Solution}
Traditional process mining techniques generally rely on a single event log to analyze processes. While this approach can be effective for isolated or straightforward workflows, it often struggles in modern organizational environments where multiple interrelated processes operate concurrently. In such contexts, analyzing a single event log can produce incomplete insights and fail to capture the complexity of cross-process interactions.
\\
\\
Moreover, preparing Multilevel processes using traditional methods is typically resource-intensive and error-prone. Manual mappings of events to cases or processes require significant setup effort, increase the risk of inconsistencies, and complicate scaling to organization-wide analyses. These challenges can slow down the delivery of actionable insights and reduce the reliability of process models.
\\
\\
Object-Centric Process Mining (OCPM) addresses these limitations by leveraging existing business data structures instead of relying on manually engineered event logs. In OCPM, each business object—such as Orders, Items, Deliveries, or Invoices—is represented as a relational table, and the natural relationships between objects automatically define process connections. Users simply select the relevant objects for analysis, and the organizational process model is constructed automatically.
\\
\\
By eliminating the need for manual log preparation and case mapping, OCPM significantly reduces configuration complexity, ensures consistent data quality, and accelerates the time to actionable insights. This approach enables comprehensive, organization-wide process visibility while minimizing effort, making it particularly suitable for large-scale, multi-process environments.
\subsubsection{Key Data Components in OCPM}
\begin{figure}
    \centering
    \includegraphics[width=1\linewidth]{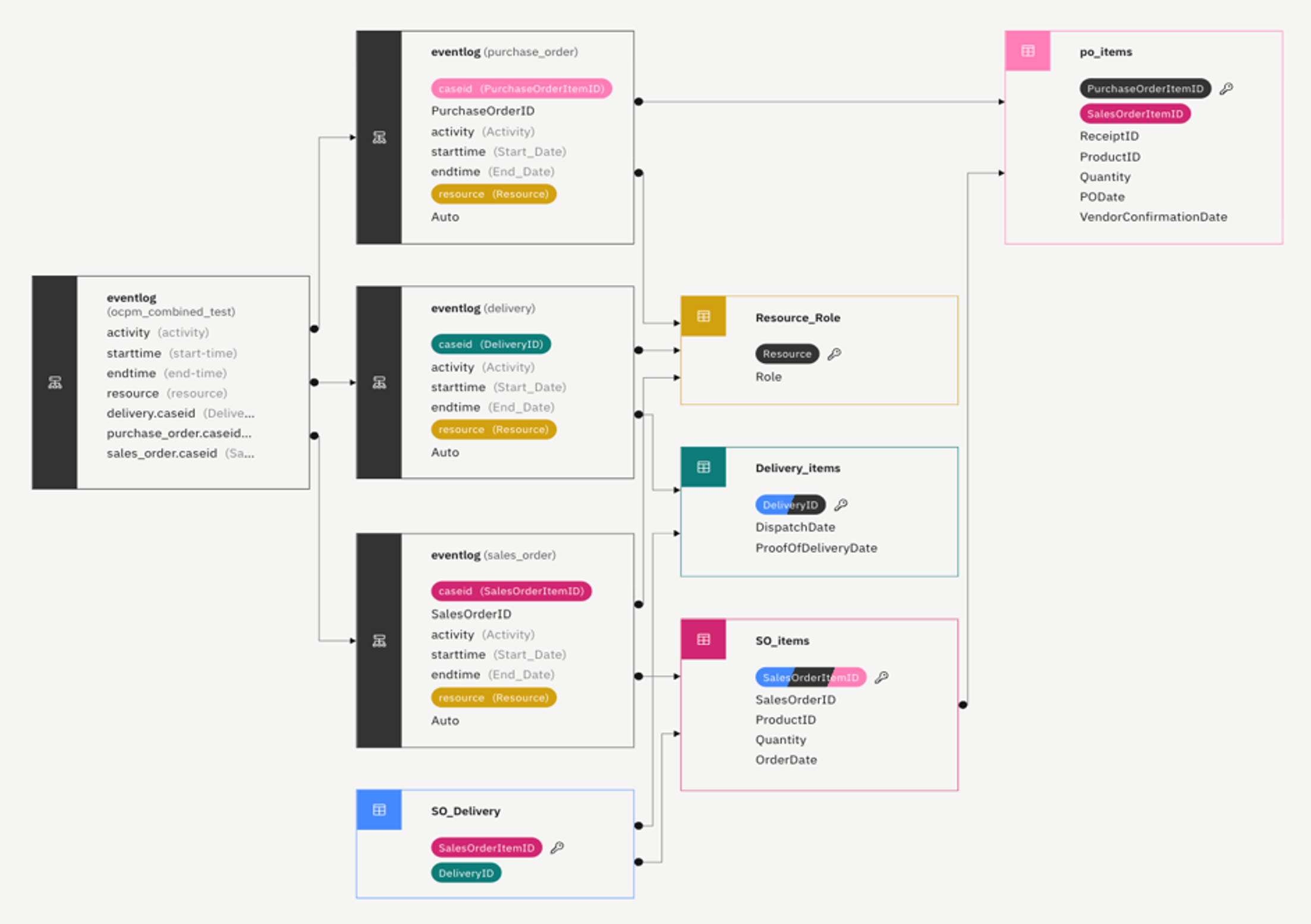}
    \caption{Illustrates the typical structure of event logs and object table joins.}
    \label{fig:Figure_ref_12}
\end{figure}

\begin{itemize}
    \item \textbf{Event Log Tables} capture the sequence of activities, timestamps, and resources involved in processes. Examples include activities such as “Order Created,” “Item Shipped,” or “Invoice Paid”.
    \item \textbf{Object Tables} represent the central business entities involved in multiple processes. Examples include Orders, Items, Deliveries, and Customers.
    \item \textbf{Connections Between Logs and Objects} each event log can be linked to one or more object tables. Shared object tables serve as bridges between different event logs, allowing for cross-process analysis and providing a complete view of organizational operations (see Figure~\ref{fig:Figure_ref_12}).
\end{itemize}
\subsubsection{Multi-Process Context \& Automatic Path Finding in OCPM}
Modern organizations typically execute multiple processes simultaneously that are often interrelated and share common business entities. In Object-Centric Process Mining (OCPM), these multiple processes coexist within the same organizational environment, and their interconnections are naturally captured through shared business objects. For instance:
\begin{itemize}
    \item \textbf{Purchase Order Process} → linked to Purchase Order Items.
    \item \textbf{Sales Order Process} → linked to Sales Items.
    \item \textbf{Delivery Process} → linked to Delivery Items.
\end{itemize}
Each of these processes maintains its own event log, documenting the activities, timestamps, and resources specific to that workflow. However, the presence of shared objects such as Items or Orders allows OCPM to establish meaningful connections across these seemingly independent processes. This inter-object connectivity enables seamless navigation across workflows and supports comprehensive, cross-process analysis, providing a holistic view of organizational operations.
\subsubsection{Automatic Path Discovery Between Processes}
\begin{figure}
    \centering
    \includegraphics[width=0.8\linewidth]{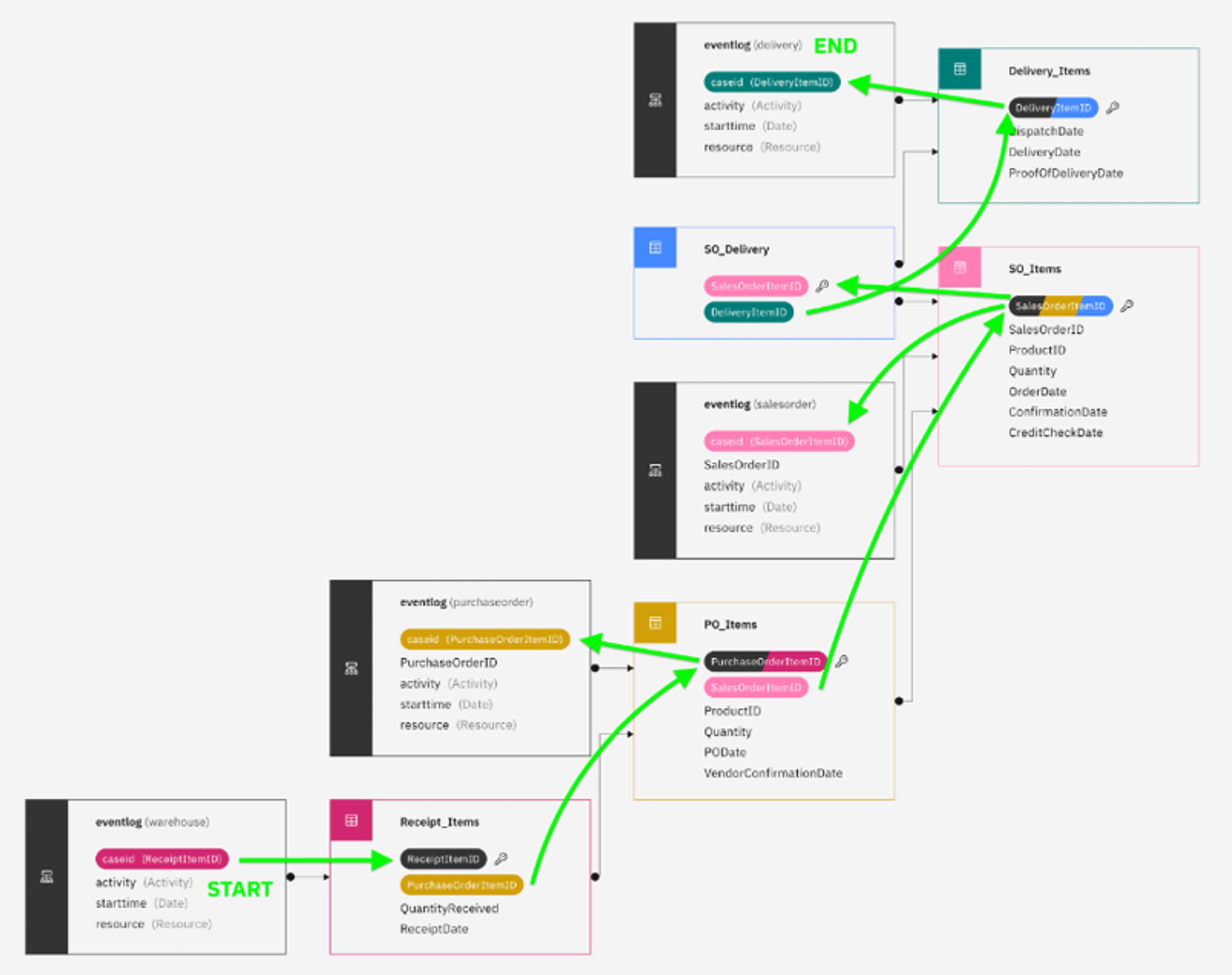}
    \caption{Illustrates an accepted path, where all relevant primary keys and case IDs are correctly connected, forming a valid and complete process chain.}
    \label{fig:Figure_ref_13}
\end{figure}

\begin{figure}
    \centering
    \includegraphics[width=0.8\linewidth]{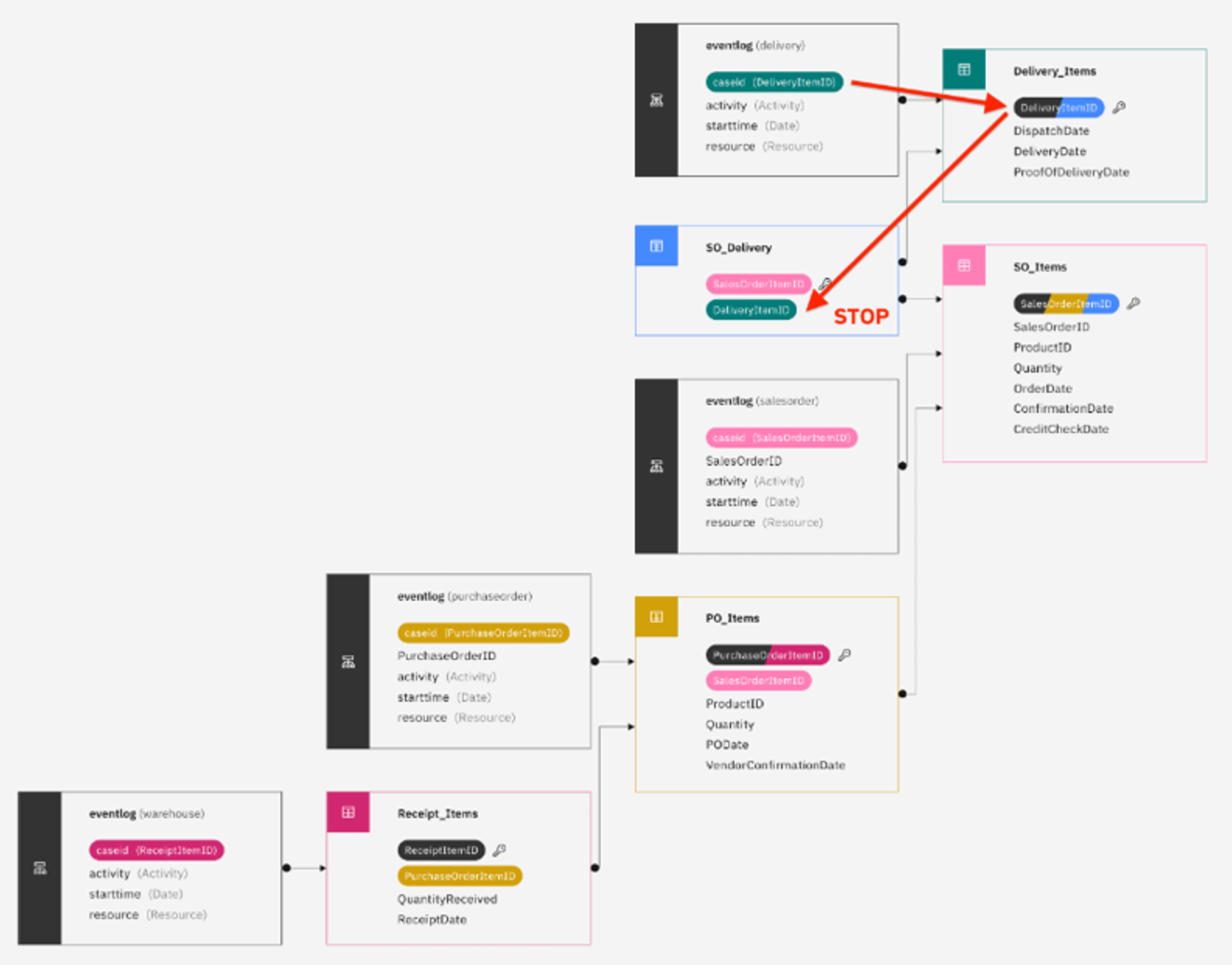}
    \caption{Depicts a discarded path, which lacks key relationships and is therefore excluded from analysis.}
    \label{fig:Figure_ref_14}
\end{figure}

One of the key innovations of OCPM is its ability to automatically discover valid connections between interrelated processes using a path-finding algorithm. The algorithm systematically evaluates potential paths between object tables and their associated event logs to determine which paths accurately reflect the underlying business relationships.
\\
\\
To ensure the integrity and completeness of the process model, the algorithm retains only those paths that satisfy two critical criteria:
\begin{enumerate}
    \item \textbf{Coverage of Primary Keys in Object Tables}: The path must include all primary keys of the involved object tables, ensuring that each business entity is correctly represented in the process chain.
    \item \textbf{Coverage of Case IDs in Event Logs}: The path must link to the appropriate case identifiers in the event logs, guaranteeing process instances traceability.
\end{enumerate}
See Figure~\ref{fig:Figure_ref_13}.
\\
\\
Paths failing to satisfy either of these criteria are automatically discarded, preventing incomplete or inconsistent process connections (see Figure~\ref{fig:Figure_ref_14}).
\\
\\
By automating path discovery, OCPM eliminates the need for manual mapping, log stitching, or custom transformations, reducing errors and accelerating the preparation and analysis of multi-process datasets. This capability is particularly valuable in large-scale environments, where the complexity and volume of interconnected processes would make manual path definition infeasible.
\subsubsection{Organizational Mining Setup}
Enabling Organizational Mining in Object-Centric Process Mining (OCPM) requires certain preconditions to ensure that the system can accurately capture and represent interconnected processes. First, multiple flat processes must exist within the same organizational environment. These processes are typically independent workflows, each with its own event log capturing activities, timestamps, and resources.
\\
\\
Second, these processes must be interconnected via shared business objects. Object tables, such as Orders, Items, or Deliveries, serve as natural bridges between event logs, allowing the system to establish meaningful relationships across processes.
\\
\\
Once these conditions are met, the “\textbf{Organizational Mining}” feature becomes available. Users can then select the processes of interest and visualize them collectively within a unified organizational model. This setup enables a comprehensive view of process interconnections, highlighting how individual workflows interact and contribute to overall operational performance. By providing a single, coherent model, the system simplifies the analysis of complex, multi-process environments while ensuring consistency and accuracy.
\subsubsection{Benefits of Organization Mining Data Preparation}
The data preparation capabilities of OCPM provide several significant advantages over traditional process mining approaches:
\begin{enumerate}
    \item \textbf{Minimal Effort}: The organizational process model is derived directly from existing business data. There is no need for manual construction of event logs, reducing setup time and human effort.
    \item \textbf{Automatic Process Definition}: The relationships between business objects automatically define process connections, eliminating the need for users to manually determine links or case hierarchies.
    \item \textbf{Elimination of Manual Mapping}: Predefined joins, custom transformations, or manual generation of case IDs are unnecessary, which significantly reduces the potential for errors and inconsistencies.
    \item \textbf{Scalability}: As new objects or processes are added to the organization, they seamlessly integrate into the existing model. This ensures that the process mining setup remains robust and adaptable to evolving business landscapes.
    \item \textbf{Enterprise-Wide Visibility}: By leveraging a single, unified data foundation, OCPM supports cross-process analysis across the entire organization. Stakeholders gain insights into both individual workflows and their interactions, enabling informed decision-making at the enterprise level.
    \item \textbf{Simplified Maintenance}: Changes to underlying systems, data schemas, or business processes automatically propagate through the organizational model. This reduces the maintenance effort required to keep the process view accurate and up-to-date.
\end{enumerate}
By streamlining the preparation of multi-process data and leveraging the inherent structure of business objects, OCPM facilitates scalable, reliable, and consistent process mining. Thus, organizations can obtain actionable insights efficiently, without the overhead and complexity typically associated with traditional approaches.
\subsection{Process Discovery \& Model Generation}
To achieve accurate process discovery within the (OCPM) framework, this approach employs a unified event log that consolidates execution data from multiple processes. A key design assumption is that each activity belongs exclusively to a single process, thereby preventing ambiguity in event-to-process mapping. This single-process activity assumption ensures clarity in identifying process boundaries and enhances the interpretability of the discovered models.
\\
\\
All event logs generated across OCPM processes are merged into a unified, time-ordered event log. This consolidated log serves as the foundation for both discovery and subsequent analytical tasks. By integrating logs at the organizational level, the approach captures the overall operational behavior, enabling a comprehensive view of interrelated workflows.
\\
\\
The temporal sequence of events is used as the guiding principle for workflow identification. The process model is derived from the chronological order of events, where start and end timestamps play a crucial role in determining activity ordering and concurrency. This allows the discovery mechanism to reconstruct workflows as they occurred in real time, capturing both sequential and overlapping executions. As a result, the unified, time-ordered event log provides a consistent and accurate representation of system behavior, supporting reliable process discovery, cross-process dependency analysis, and performance evaluation within OCPM.
\section{Model Generation - Interleaving Workflows \& Enhanced Visualization}
\begin{figure}[H]
    \centering
    \includegraphics[width=0.9\linewidth]{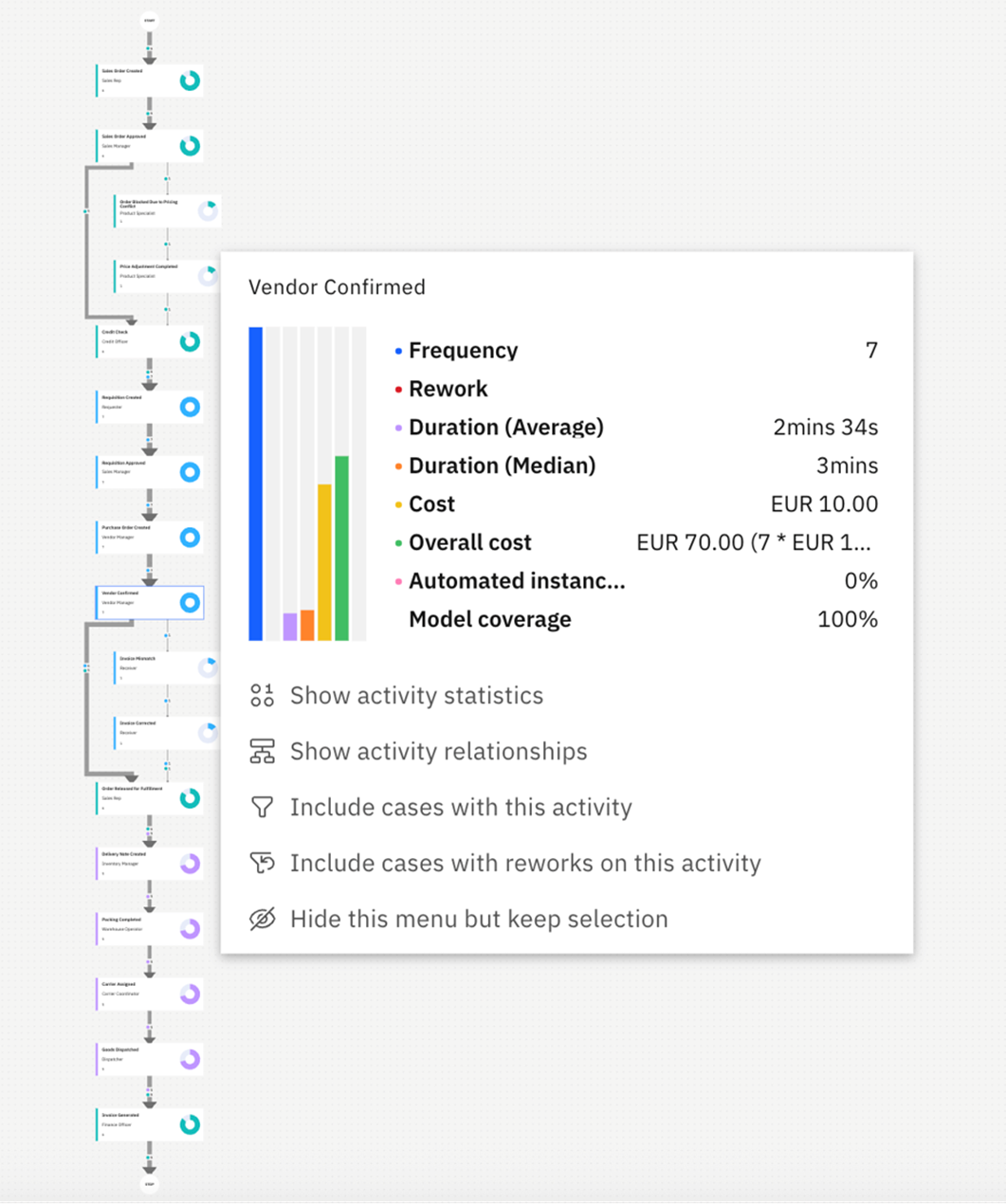}
    \caption{Illustrates the generated process model, where workflows from multiple processes are interleaved to reveal their temporal interactions and dependencies. Color-coded nodes and edges distinguish different activity types and process streams, enhancing visual clarity. Additionally, embedded performance metrics, such as activity frequency and execution duration, provide deeper insights into process behavior, execution trends, and bottleneck identification.}
    \label{fig:Figure_ref_15}
\end{figure}

Following the creation of the unified event log, the \textbf{model generation} phase constructs a comprehensive process model by \textbf{interleaving individual workflows} in temporal order. This approach captures the dynamic interactions and overlaps among processes, enabling the visualization of how multiple workflows coexist and influence each other over time.
\\
\\
The resulting model adopts an \textbf{object-centric perspective}, allowing simultaneous tracking of multiple object types such as orders, customers, and resources. This contrasts with traditional activity-centric process models by providing a richer and more contextualized representation of organizational behavior. Through this multi-object linkage, dependencies and correlations between activities across different domains can be observed and analyzed more effectively.
\\
\\
To enhance interpretability, the visualization integrates \textbf{color coding} and \textbf{styling mechanisms}. Nodes and edges are differentiated using distinct colors to represent processes, objects, or event types, improving clarity in complex model structures.
\\
\\
Furthermore, supplementary information fields, including frequency and performance metrics, are embedded within the visualization. These features allow for in-depth analysis of aspects such as activity frequency, bottlenecks, or throughput times.
\\
\\
The outcome is a \textbf{unified, object-centric process model} that provides both a high-level overview and detailed analytical capability. By interleaving workflows and enriching visualization, the model supports more accurate interpretation of process interactions, enabling better decision-making and performance optimization within the OCPM environment (see Figure~\ref{fig:Figure_ref_15}).
\newpage
\subsection{Model Statistics}
Model Statistics in Organizational Mining provide a cross-object view of process performance, enabling accurate and comprehensive analysis across organizational workflows. Unlike traditional case-based process mining, which focuses on a single case entity, Organizational Mining integrates multiple related processes—such as sales orders, purchase orders, and deliveries—through shared object relationships. The IBM approach ensures that statistical measures remain representative even within complex, multi-object interactions, delivering organization-wide insights and a true reflection of operational behavior.
\subsection{Accurate Statistics Across Processes}
\begin{figure}[H]
    \centering
    \includegraphics[width=0.8\linewidth]{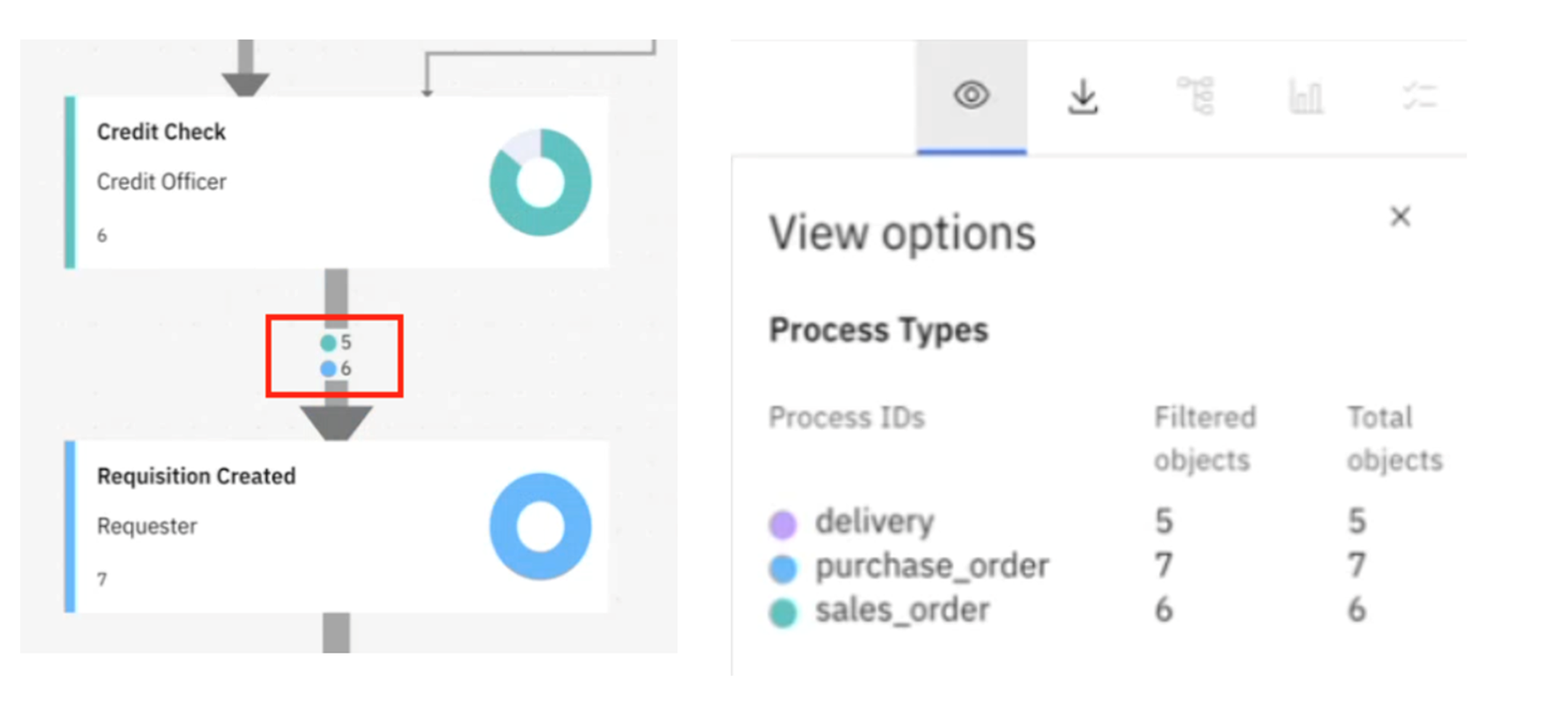}
    \caption{Illustrates the process model view showing multiple interconnected process types—delivery, purchase order, and sales order. Each color represents a distinct process type, and the counts (5, 6, 7) indicate the number of corresponding object instances contributing to the displayed path between activities such as "Credit Check" and "Requisition Created". The visual overlay highlights how object-level data is aggregated to compute shared process behavior across related entities.}
    \label{fig:Figure_ref_16}
\end{figure}

To maintain the reliability of quantitative measures, the model applies built-in strategies to mitigate statistical distortion caused by divergence and convergence. Divergence occurs when a single object instance, such as a sales order, is associated with multiple related entities (for example, several purchase orders). Conversely, convergence arises when multiple objects reference a single entity. Without corrective mechanisms, such relationships can produce skewed values for metrics such as frequency or duration.
\\
\\
By compensating for these structural variations, this approach ensures that computed metrics, including frequency and duration, remain consistent across process variations involving multiple object linkages. Within the model visualization (Figure~\ref{fig:Figure_ref_16}), these adjustments guarantee that visual indicators, such as path thickness and activity frequency counts, represent the actual operational behavior of the process, eliminating distortions introduced by divergent or convergent data structures.
\subsection{Cross-Object Computation}
\begin{figure}
    \centering
    \includegraphics[width=1\linewidth]{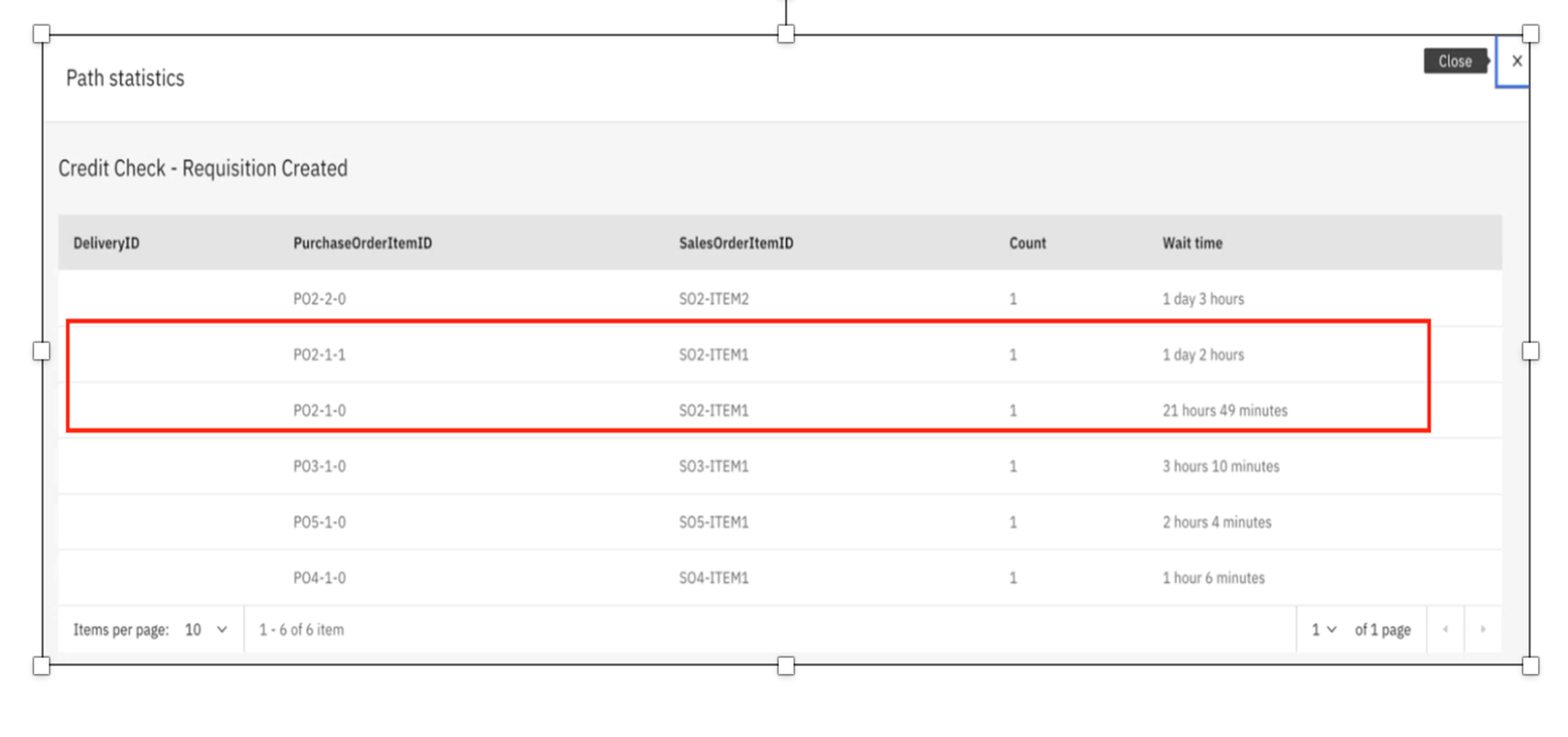}
    \caption{Shows the path-level statistics derived from object relationships. The table lists delivery identifiers, purchase order items, and sales order items linked to a selected connection, along with computed wait times. These statistics demonstrate cross-object computation for accurate performance indicators across divergent and convergent relationships.}
    \label{fig:Figure_ref_17}
\end{figure}

Organizational Mining introduces cross-object computation, where statistics are derived across multiple interconnected object types instead of single-case data. This allows the analysis of relationships that span multiple process types, such as a sales order connected to purchase orders and delivery records.
\\
\\
These computations are used in both:
\begin{itemize}
    \item Activity Statistics, which display aggregated metrics (frequency, rework, duration, cost) for each activity across linked objects;
    \item Path Statistics, which measure the relationships between activities across objects, revealing cross-process dependencies and waiting times.
\end{itemize}
In the example shown in Figure~\ref{fig:Figure_ref_17}, one sales order item is linked to multiple purchase order items. The Path Statistics table below highlights the connection between the activities "Credit Check" and "Requisition Created", showing the corresponding Delivery IDs, Purchase Order Item IDs, and Order Item IDs along Sales with count and wait time. This demonstrates how the model accurately captures cross-object process relationships and aggregates time and frequency metrics across related process flows.
\subsection{Key Measures}
The model provides a detailed statistical breakdown for each activity. By selecting an activity node within the process model, users can view the following measures directly:
\begin{itemize}
    \item \textbf{Frequency}: Number of times the activity appears in the process.
    \item \textbf{Rework}: Number of times the same activity repeats within the same case.
    \item \textbf{Duration (Average / Median)}: Typical time taken to complete the activity.
    \item \textbf{Cost / Overall Cost}: The individual and cumulative cost associated with the activity.
    \item \textbf{Automated Instance Ratio}: The percentage of activity executions that are automated.
    \item \textbf{Model Coverage}: The extent to which relationships involving the activity are visualized in the process model (a 100\% coverage indicates all relationships are visible).
\end{itemize}
\subsection{Insights}
By combining divergence-aware computation, cross-object statistics, and granular activity-level metrics, the IBM Object-Centric Process Mining model delivers a comprehensive and organization-wide statistical view of processes. It provides an overall understanding of how different business entities interact without relying solely on case-level data and aggregates event counts and durations to reflect true end-to-end performance.
\\
\\
This enables organizations to identify inefficiencies, detect bottlenecks, and derive actionable insights across interconnected process dimensions.
\subsection{Throughput Time Between Activities}
Throughput Time is one of the most fundamental metrics in process analysis, it represents the total elapsed time between two activities or milestones in a process. In traditional, case-based process mining, this is calculated within the boundaries of a single case or instance (for example, one purchase order or one customer request).
\\
\\
However, this case-based view often oversimplifies real-world processes, where multiple business objects interact with each other, such as requisitions, purchase orders, invoices, or deliveries. When such interlinked activities are forced into a flat, case-based model, distortions occur due to event divergence (one activity leading to many) and event convergence (many activities merging into one).
\\
\\
Organizational Mining Throughput Time addresses this limitation by measuring execution durations across related process objects, rather than within isolated cases. It provides a more accurate and realistic representation of process performance, ensuring that timing metrics remain reliable even in complex, interconnected process networks.
\subsection{Throughput Time in OCPM}
In the Object-Centric Process Mining (OCPM) model, each event can be associated with multiple process objects simultaneously. For example, a single “Approval Completed” event may be linked to both a Requisition and a Purchase Order.
\\
\\
Because of this, OCPM does not rely on traditional, linear case identifiers. Instead, it traces the relationships between objects to compute throughput time, effectively following the actual flow of business logic as it moves through different object types.
\\
\\
This computation model ensures that:
\begin{itemize}
    \item Both start and end activity objects are evaluated within the same logical interval. Throughput time reflects the duration between truly related events (e.g., a requisition that directly results in a purchase order), rather than random or unrelated timestamps.
    \item Skew from event divergence or convergence is avoided. OCPM accounts for real-world patterns such as one-to-many and many-to-one relationships between activities, removing artificial inflation or compression of time measurements that occur in flat models.
\end{itemize}
In essence, OCPM throughput time captures how long processes actually take across object interactions, not just within isolated, artificial cases.
\\
\\
Example: Multiple Paths Between the Same Events.
\\
\\
Consider a scenario with multiple possible paths connecting the same start and end events:
\begin{itemize}
    \item \textbf{Requisition Created → Purchase Order Created} 
\end{itemize}
\begin{figure}
    \centering
    \includegraphics[width=1\linewidth]{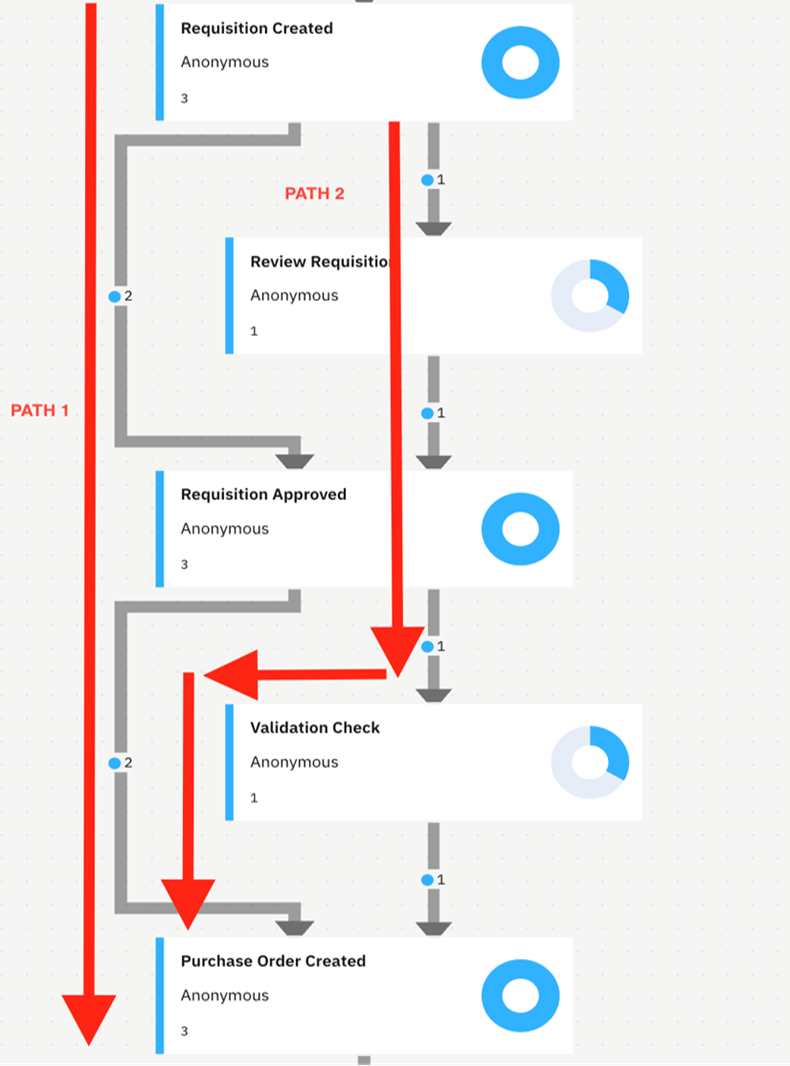}
    \caption{Represents multiple paths between the same start and end events, which OCPM treats as a single logical flow when calculating metrics like average throughput time, regardless of the intermediate steps.}
    \label{fig:Figure_ref_18}
\end{figure}
In the process diagram Figure~\ref{fig:Figure_ref_18}, two possible routes exist:
\begin{itemize}
    \item \textbf{Path 1}: "Requisition Created" → "Requisition Approved" → "Purchase Order Created".
    \item \textbf{Path 2}: "Requisition Created" → "Review Requisition" → "Requisition Approved" → "Purchase Order Created".
\end{itemize}
Although the intermediate steps differ, both paths ultimately represent the same start-to-end relationship between “Requisition Created” and “Purchase Order Created.”
\\
\\
When calculating statistics such as average throughput time, OCPM treats this event pair as a single logical measurement.
\\
\\
This means:
\begin{itemize}
    \item The metric depends only on the absolute time difference between:
    \begin{itemize}
        \item the start time of the “From” activity (Requisition Created);
        \item the end time of the “To” activity (Purchase Order Created).
    \item Intermediate variations, such as whether the case passes through "Review Requisition" or "Validation Check", do not affect the computed throughput time. 
    \end{itemize}
\end{itemize}
By abstracting away from the specific path taken, OCPM ensures that throughput time analysis remains path-independent and universal.
\\
\\
The focus stays on \textbf{actual elapsed time} between key boundary events, providing a consistent and reliable performance measure across all process variants.
\subsection{Why This Matters}
\begin{itemize}
    \item \textbf{Accurate Duration Measurement}: Reflects real-world inter-object interactions instead of oversimplified single-case sequences.
    \item \textbf{Path Independence}: Eliminates the need to treat every route separately; ensures consistency in metrics.
    \item \textbf{Process Reliability}: Provides insights that hold true across departments, systems, and object relationships.
    \item \textbf{Organizational Insight}: Enables global visibility into how long processes truly take from initiation to completion — regardless of complexity.
\end{itemize}
\subsection{Conformance}
Conformance checking is natively performed through the use of ad-hoc trace filters, custom metrics, and dedicated analytics dashboards, enabling a comprehensive analysis of deviations from the expected process model.
\\
\\
The \textbf{trace filters} allow the detection of deviation by including activities that should not appear in the workflow of the process or by using process flow filters that define specific sequences of activities to be monitored.
\\
\\
\textbf{Custom metrics} enable the creation of targeted Key Performance Indicators (KPIs) to assess whether certain measures, such as the throughput time between activities belonging to different processes, are outside the expected range.
\\
\\
Finally, the \textbf{analytics dashboards} provide a consolidated view of the overall conformance status through configurable widgets that combine user-defined queries, filters, and custom metrics, thereby supporting dynamic and adaptable monitoring according to specific analytical needs.
\section{Conclusions}
With Organization Mining, IBM Process Mining 2.0 evolves into a strategic tool for process optimization, data-driven insight generation, and organizational improvement. Built on the foundations of Object-Centric Process Mining, Organization Mining introduces a significant advancement in process analysis, centered around its greatest strength: the simplified data preparation. Analysts can start from existing flat projects without the need for complex mappings or extensive engineering efforts, resulting in a setup that is fast, intuitive, and easily accessible.
\\
\\
This simplified data preparation provides several key advantages, including easier integration of multiple processes across the organization, accurate and distortion-free statistical analyses, faster adoption and simplified maintenance, and scalability to accommodate even the most complex enterprise environments. Consequently, Organization Mining enables a higher level of analytical insight, allowing processes to be examined not in isolation but in relation to one another and to the resources involved. This interconnected perspective empowers faster, more informed decision-making, enhances organizational performance, and supports continuous improvement and evolution.

\bibliographystyle{unsrtnat}
\bibliography{references}
\end{document}